\documentclass{aa}  

\usepackage{natbib}
\usepackage{graphicx}	
\graphicspath{{figures/}}	
\usepackage{amsmath}	
\usepackage{amsfonts}
\usepackage{amssymb}
\usepackage{txfonts}
\usepackage{hyperref}
\hypersetup{
  colorlinks=true, 
  urlcolor=blue, 
  linkcolor=blue, 
  citecolor=blue 
}

\usepackage{xcolor}
\def\nh{N_{\rm H}}
\def\dlg{\Delta\log\varg}
\def\Teff{T_{\rm eff}}
\def\chandra{\textit{Chandra}}
\def\xmm{\textit{XMM-Newton}}

\begin{document} 

    \title{Examining the evolution of the supersoft X-ray source RX~J$0513.9-6951$}

    \author{A.~Tavleev \and V.~F.~Suleimanov \and K.~Werner \and A.~Santangelo}

    \institute{Institut f\"ur Astronomie und Astrophysik, Kepler Center for Astro and Particle Physics, Universit\"at T\"ubingen, Sand 1, 72076 T\"ubingen, Germany \\ \email{tavleev@astro.uni-tuebingen.de}}

    \date{Received xxx / accepted xxx}
 
  \abstract
   {Supersoft X-ray sources~(SSSs) are thought to be accreting white dwarfs~(WDs) in close binary systems, with thermonuclear burning on their surfaces. The SSS RX~J$0513.9-6951$ in the Large Magellanic Cloud~(LMC) exhibits cyclic variations between optical low and high states, which are anti-correlated with its X-ray flux. This behaviour is believed to be the result of the periodic expansion and contraction of the~WD due to variations in the accretion rate in the system.}
   {We analyse the eight high-resolution~XMM and six grating \textit{Chandra} spectra of RX~J$0513.9-6951$ with our grid of model atmosphere spectra of hot WDs computed under the assumption of local thermodynamic equilibrium. Our aim is to test a contraction model of the source variability by tracking the evolution of the WD properties.}
   {We use a recently computed grid of hot WD~model atmospheres, spanning a wide range of effective temperatures~($\Teff=100-1000\,\rm kK$ in steps of $25\,\rm kK$) and eight values of surface gravity. The LMC~chemical composition of the atmospheres was assumed.}
   {The obtained fitting parameters~(effective temperature $\Teff$, surface gravity $\log\varg$, and bolometric luminosity~$L$) evolve on the $\Teff - \log\varg$ and $\Teff - L$ planes. This evolution follows the model tracks of WDs with masses of $1.05-1.15\,M_{\sun}$ and thermonuclear burning on the surface. We show that, when the source has a relatively small photospheric radius and is optically bright, it lies below the stable-burning strip with a relatively low bolometric luminosity. Conversely, the fainter optical states correspond to higher bolometric luminosity and larger photospheric radii of the hot~WD. RXJ0513 lies within the stable-burning strip during this state. This means that the optical brightness of the system is lower when the~WD is larger, more luminous, and illuminates the accretion disc more effectively. These results contradict the contraction model, which predicts the opposite behaviour of the source. We use a model that assumes that the far UV/soft X-ray flux is reprocessed into the optical band due to multiple scattering in the cloud system above the accretion disc. More significant illumination can lead to rarefying of the cloud slab, thereby reducing the reprocessing efficiency and making the source fainter in the optical band.}
   {}

   \authorrunning{Tavleev et al.}
   \titlerunning{Supersoft X-ray Source RX~J$0513.9-6951$}

   \maketitle
%

\section{Introduction}

Supersoft X-ray sources~(SSSs) represent a sub-class of cataclysmic variable~(CV) stars characterised by a high mass-accretion rate of~${\sim}10^{-7}\,M_\odot\, \rm yr^{-1}$ and (quasi-)steady state thermonuclear burning on the white dwarf~(WD) surface~\citep{van_den_Heuvel_etal1992, Rappaport_etal1994}. CVs are close binary systems composed of a WD primary and a main-sequence or red (sub-)giant companion that fills its Roche lobe. Consequently, matter transfers from the companion to the~WD, potentially forming an accretion disc.

The first~SSSs were identified in the Magellanic Clouds by the \textit{Einstein} observatory~\citep{Long_etal1981, Seward_Mitchell1981}, and they were later classified as a new type of X-ray source during ROSAT~observations~\citep{Trumper_etal1991}. These sources are defined by their extremely soft thermal X-ray spectra, with blackbody temperatures ranging from $15-80$~eV ($150-900\,\rm kK$) and high luminosities of~${\sim} 10^{36}-10^{38} \rm\, erg/s$, which are comparable to the Eddington luminosity of a solar mass object~\citep{Kahabka_van_den_Heuvel1997, Greiner2000}. To date, more than $100$~SSSs have been identified in approximately $20$~external galaxies, the Magellanic Clouds, and our Galaxy~\citep{Greiner2000, Kahabka2006, Maitra_etal2022}.

The luminous transient soft X-ray source RX~J$0513.9-6951$~(hereafter RXJ0513) in the Large Magellanic Cloud (LMC) was discovered in the ROSAT All-Sky Survey~\citep{Schaeidt_etal1993, Pakull_etal1993, Cowley_etal1993}. Optical spectroscopic analysis revealed strong emission lines of \element{H}, \ion{He}{II} and several higher ionisation emission features with broad wings, indicating the presence of an accretion disc, while the accretor is most likely a~WD~\citep{Crampton_etal1996, Southwell_etal1996}. Thanks to the Massive Compact Halo Objects project (MACHO) survey~\citep{Alcock_etal1996}, a~$3.5$~yr light curve of the V~${\sim}\, 17^m$~(absolute $M_V\,{\sim}-1.5$) optical counterpart of this source was obtained, which revealed recurrent low states~(dropping by ${\sim}\, 1^m$), which last $20-40$~days and repeat every $100-200$ days. Moreover, these optical low states are accompanied by X-ray outbursts, so the optical and X-ray states are strictly anti-correlated~\citep{Southwell_etal1996}. Such variability is explained by the so-called contraction model~\citep[e.g.][]{Reinsch_etal1996, Reinsch_etal2000, Southwell_etal1996, Hachisu_etal2003a, Hachisu_etal2003b}, which suggests that the observed changes occur due to a change in the mass accretion rate, and therefore, a change in the intensity of thermonuclear burning on the WD~surface. High rate of thermonuclear energy release leads to a significant expansion of a WD~photosphere, and the peak in emergent radiation lies outside the observed X-ray band, which makes the source faint in X-ray. At the same time the bolometric luminosity of the expanded~WD is high, and it effectively illuminates the accretion disc, resulting in high optical flux due to the reprocessing of the WD~radiation. This scenario corresponds to the optically bright and X-ray faint state. On the other hand, a low mass accretion rate causes the~WD to shrink, and its outgoing radiation is observed in the soft X-ray band. However, a smaller photospheric radius and decreased WD~luminosity make the accretion disc illumination less significant, resulting in faint optical brightness. The origin of the periodicity in the mass accretion rate changes is not completely clear.

\citet{Reinsch_etal2000} proposed a limit-cycle model, where periodic changes in the accretion disc viscosity, driven by irradiation from the~WD, could lead to variations in the accretion rate and subsequent changes of the~WD~photosphere size. In this model, the mass transfer from the companion star to the WD remains constant. Alternatively, \citet{Hachisu_etal2003a, Hachisu_etal2003b} suggested the `accretion wind evolution' model, in which a self-sustained optically thick wind from the~WD impacts the companion star, reducing the mass transfer rate. This diminishes the wind, causing the mass transfer rate to increase again. The behaviour of such a wind modulates the accretion rate onto the~WD, resulting in the transitions between high and low optical states.
\citet{Zhao_etal2022} tested the contraction model by performing a~\textsc{MESA} simulation of the WD~accretion process and its subsequent evolution in a binary star with a periodic mass transfer in order to reproduce the observed optical variability of RXJ0513. A WD~irradiation of the companion star was proposed to be an origin of the periodic mass transfer.

\citet{McGowan_etal2005} confirmed the contraction model by observing~RXJ0513 with the \xmm~telescope to obtain high-resolution spectra of the object and track the X-ray evolution of the WD. However, interpreting SSS~spectra presents significant challenges. While low-resolution spectra can often be described by a blackbody model~\citep[as done by][]{McGowan_etal2005}, this approach lacks physical accuracy. In general, blackbody fitting leads to an overestimation of the interstellar neutral hydrogen column density,~$\nh$, and underestimation of the temperature, resulting in an overestimated luminosity~\citep[see e.g.][]{Greiner_etal1991, Heise_etal1994}. High-resolution spectra of SSSs are considerably more complex, featuring numerous emission and absorption lines. 

The spectra of~SSSs can be more accurately described by theoretical spectra of hot WD~atmospheres. This approach has been applied by numerous authors using both local thermodynamic equilibrium~\citep[LTE, e.g.][]{Heise_etal1994, Ibragimov_etal2003, Suleimanov_Igragimov2003, Burwitz_etal2007, Swartz_etal2002}, and non-LTE~\citep[e.g.][]{HartmannHeise1996, Lanz_etal2005, Petz_etal2005, vanRossum2012, Rauch_etal2010} assumptions. 
Non-LTE effects are non-negligible, and non-LTE models are more physically accurate, all else being equal. However, the physical conditions of the radiating regions in~SSSs are not entirely clear and may differ significantly from the simplest plane-parallel hydrostatic atmosphere for which non-LTE effects are considered. The uncertainties associated with factors like radiation-driven winds, boundary layers, accretion disc irradiance are likely much more significant than non-LTE effects. Therefore, we believe that LTE~models provide comparable accuracy~\citep[see][]{Suleimanov_etal2024, Tavleev_etal2024} and can be used to analyse the soft X-ray spectra of~SSSs. 
Moreover, LTE~computations are much faster, allowing for the calculation of more extensive model grids and the inclusion of a larger number of ions, their excited levels, and a greater number of spectral lines compared to the non-LTE case.

Recently \citet{Suleimanov_etal2024} further refined the LTE~approach, including a more extensive set of spectral lines and photoionisation opacities from the excited levels of heavy element ions. They successfully apply a computed set of model atmosphere spectra to the interpretation of grating \chandra~and \xmm~spectra of two SSS~sources, CAL~$83$ and RXJ0513; for latter source only one \chandra~observation was used. \citet{Tavleev_etal2024} utilised the code by~\citet{Suleimanov_etal2024} to compute a specific grid of LTE~models for analysing the~eROSITA and \xmm~spectra of the nova AT~2018bej, focusing specifically on the chemical composition. It was concluded that LTE~model atmospheres can be used to analyse the available X-ray spectra of classical novae during their SSS~state.

In this work we analyse the high-resolution~XMM and \chandra~spectra of 
RXJ0513 using the LTE~atmosphere grid calculated by \citet{Suleimanov_etal2024}. Our aim is to test the contraction model of the source's variability by tracking the evolution of the WD~parameters and comparing them with the optical brightness of the source. The X-ray observations used are described in Sect.~\ref{sect:Observations}. We describe the used models atmospheres and the fitting technique in Sect.~\ref{sect:Method}. The results of spectral fitting are presented in Sect.~\ref{sect:Results}, while in Sect.~\ref{sect:Discussion} we discuss the possible nature of the obtained evolutionary trends. We conclude and summarise our findings in Sect.~\ref{sect:Conclusions}.


\section{Observations}
\label{sect:Observations}

The source was observed by the \textit{Chandra} X-ray observatory using the High Resolution Camera~(HRC-S) and the Low Energy Transmission Grating~(LETG). A total of six spectra are publicly available in the Chandra Grating-Data Archive and Catalog\footnote{\url{http://tgcat.mit.edu}}~\citep[TGCat, ][]{Huenemoerder_etal2011}, the {\it Chandra} observation log is presented in~Tab.~\ref{tab:log_chandra}. Below we refer to the observation with ObsID~$3503$ as the C0~spectrum, while the spectra with ObsIDs~$5440-5444$ are labelled $\rm C1-C5$. To increase the signal-to-noise ratio the positive and negative first-order spectra were co-added using the \texttt{combine\_grating\_spectra} task of the Chandra Interactive Analysis of Observations~\citep[CIAO,][]{Fruscione_etal2006, CIAO_ascl} package. For the fitting purposes the spectra were rebinned to at least $30$~counts per bin in the $0.21-0.65\,\rm keV$~energy range. 

\citet{McGowan_etal2005} published the analysis of \xmm~observations of the source. The high-resolution spectra, obtained using the Reflection Grating Spectrometer~\citep[RGS,][]{denHerder_etal2001} instrument, are available for eight observations~(see Tab.~\ref{tab:log_xmm}). It should be noted that the observation with ObsID~$0151410101$ was affected by high solar flaring activity and contains no useful data. All spectra are available in the \xmm~Science Archive\footnote{\url{http://nxsa.esac.esa.int/nxsa-web}}, only the first-order spectrum was used without any rebinning. We did not combine the data from~RGS1 and~RGS2 but fitted them simultaneously. Below we refer to these spectra as $\rm X1-X8$.


{\setlength{\tabcolsep}{4pt}  
\begin{table}
    \caption{\textit{Chandra} HRC-S/LETG observation log of RXJ0513.}
\begin{center}
    \begin{tabular}{ccccc}\hline\hline \noalign{\smallskip}
        & ObsID & Date & MJD & Exp. time, ks \\[0.3em] \hline\noalign{\smallskip}
        C0 & $3503$ & 2003 December 24 & $52997.29$ &  $47.65$  \\ [0.3em]
        C1 & $5440$ & 2005 April 20 & $53480.80$ &  $24.53$     \\ [0.3em]
        C2 & $5441$ & 2005 April 27 & $53487.95$ &  $25.00$     \\ [0.3em]
        C3 & $5442$ & 2005 May 03 & $53493.24$ &  $25.50$       \\ [0.3em]
        C4 & $5443$ & 2005 May 13 & $53503.82$ &  $22.48$       \\ [0.3em]
        C5 & $5444$ & 2005 May 19 & $53509.09$ &  $24.99$       \\
        \hline 
    \end{tabular}
\end{center}
\label{tab:log_chandra}
\end{table}

\begin{table}
    \caption{\xmm~observation log of RXJ0513.}
\begin{center}
    \begin{tabular}{ccccc}\hline\hline \noalign{\smallskip}
        & ObsID & Date & MJD & Exp. time$^a$ \\[0.3em] \hline\noalign{\smallskip}
           & $0151410101$ & 2004 April 28 & $53123.74$ & $0.0^b$ \\ [0.3em]
        X1 & $0151412101$ & 2004 May 02 & $53127.08$ & $16.84/16.83$ \\ [0.3em]
        X2 & $0151412201$ & 2004 May 05 & $53130.81$ & $17.72/17.70$ \\ [0.3em]
        X3 & $0151412301$ & 2004 May 09 & $53134.91$ & $17.64/17.61$ \\ [0.3em]
        X4 & $0151412401$ & 2004 May 12 & $53137.69$ & $25.62/25.62$ \\ [0.3em]
        X5 & $0151412501$ & 2004 May 16 & $53141.06$ & $13.84/13.82$ \\ [0.3em]
        X6 & $0151412601$ & 2004 May 18 & $53143.35$ & $13.84/13.83$ \\ [0.3em]
        X7 & $0151412701$ & 2004 May 26 & $53151.24$ & $18.03/18.03$ \\ [0.3em]
        X8 & $0151412801$ & 2004 May 28 & $53153.27$ & $15.83/15.83$ \\
        \hline 
    \end{tabular}
\end{center}
{\small Notes: (a)~-- Exposure time in ks for~RGS1 and RGS2~instrument, respectively; (b)~-- the first observation has very high background and does not contain useful data.}
\label{tab:log_xmm}
\end{table}
}

\section{Method}
\label{sect:Method}

A new grid of hot LTE~model atmospheres, recently presented by \citet{Suleimanov_etal2024}, was used to analyse the observed spectra. These model spectra were successfully used for the analysis of the classical SSS~CAL~$83$ and one of the \textit{Chandra} spectra of RXJ0513~(spectrum~C3).

The models were computed using a code based on Kurucz's \textsc{atlas}~\citep{Kurucz1970}, which was modified for high temperatures and high densities \citep{Ibragimov_etal2003, Suleimanov_etal2006, Suleimanov_etal2015}. The models account for the$15$~most abundant chemical elements from~\element{H} to~\element{Ni} and about $20,000$~lines from CHIANTI, Version~3.0, atomic database~\citep{Dere_etal1997}. The main parameters of the model are the effective temperature $\Teff$ and the gravity parameter $\dlg = \log\varg - \log\varg_{\rm Edd}$, which indicates the distance of model from the Eddington limit~\citep[see details in][]{Suleimanov_etal2024}:
\begin{equation}
    \log\varg_{\rm Edd} = \log(\sigma_{\rm e}\sigma_{\rm SB}\Teff^4 /c) = 4.818 + 4\log(\Teff/10^5 \rm K),
\end{equation}
were~$\sigma_{\rm e},\,\sigma_{\rm SB}$~and~$c$ are the Thomson cross-section for the electron, Stefan–Boltzmann constant, and the speed of light, respectively.

The grid was computed for~$\Teff$ in the range of~$100-1000\rm\, kK$ in steps of~$25\,\rm kK$. The $\dlg$~parameter has eight values: $0.1, 0.2, 0.4, 0.6, 1.0, 1.4, 1.8$~and~$2.2$. The grid was calculated for three chemical compositions; we used the LMC~composition, where the hydrogen-helium mix is set to solar, while the abundance of heavy elements is set to half-solar~$(A = 0.5)$. The model grid is available\footnote{\url{https://github.com/HEASARC/xspec_localmodels/tree/master/sss_atm}} as an XSPEC\footnote{\url{https://heasarc.gsfc.nasa.gov/docs/xanadu/xspec/}}~\citep{Arnaud1996, XSPEC_ascl} table.

All the observed X-ray spectra were fitted using the models described above. We used the WD~mass~$M$ and radius~$R$ as free parameters instead of~$\dlg$ and normalisation parameter. A uniform prior distribution was set for $\Teff$, $M$ (in range $0.3-1.4\,\rm M_{\sun}$), and $R$ (in range $(2-20) \times 10^8\,\rm cm$). We note that the strict upper limit was set for the WD~mass. Another limitation applied is based on the fact that the WD~radius must be greater than the cold WD~radius~$R_{\rm cold}$ at such a mass~\citep{Nauenberg1972}. The Tübingen-Boulder ISM model \texttt{tbabs} \citep{Wilms2000} was employed to account for absorption, with a uniform prior distribution set for the hydrogen column density $\nh$~in the range of $(1-10)\times 10^{20} \,\rm cm^{-2}$.

To find the best-fitting parameters and estimate their uncertainties, we used the following technique. First, we employed the Bayesian approach. The analysis was conducted using the software Bayesian X-ray Analysis~\citep[BXA,][]{Buchner_etal2014, BXA_ascl}, which connects the nested sampling package UltraNest\footnote{\url{https://johannesbuchner.github.io/UltraNest/}}~\citep[][]{Buchner2021, UltraNest_ascl} with~\textsc{xspec}\,\footnote{\url{https://heasarc.gsfc.nasa.gov/docs/xanadu/xspec/}}\citep{Arnaud1996, XSPEC_ascl}. The nested sampling Monte Carlo algorithm MLFriends~\citep{Buchner_etal2014, Buchner2019} was used to obtain posterior probability distributions and the Bayesian evidence. 

The usage of C-statistics~\citep{Cash1979} is preferable instead of the $\chi^2$-statistics for low-count spectra. We used its~\textsc{xspec} implementation~\texttt{cstat} as a likelihood in order to determine the best-fit parameters. The parameter uncertainties were derived from the~$0.16$~and~$0.84$~quantiles of the posterior distribution (which corresponds to $68\%$ confidence level, CL). 

However, we consider these statistical errors to be significantly underestimated. To obtain more realistic uncertainties, we added a systematic error to the spectra such that the reduced~$\chi^2_\nu$ approached unity, after which we re-estimated the parameter confidence intervals. We then combined the statistical confidence intervals (derived using~\texttt{cstat}) with the systematic ones (calculated using~$\chi^2$). The final parameter value was taken as the mean of the combined confidence interval.

Due to such a technique of estimating the model parameters and its confidence intervals in the tables below we do not provide the statistics value or the goodness-of-fit estimating criterion~(like $\chi^2_{\nu}$).

\section{Results}
\label{sect:Results}
\subsection{Hydrogen column density}

First, we fitted the observed spectra using the interstellar absorption column density $\nh$ as a free parameter. It is found that the absorption parameter obtained differs significantly across the various spectra. For the \xmm~spectra, $\nh$ is contained within the range $(1-3.5)\times 10^{20}\,\rm cm^{-2}$, while for the \chandra~spectra $\nh$ lies within $(3.7-5.4)\times 10^{20}\,\rm cm^{-2}$. All obtained values are lower than the average Galactic absorption column in the direction of the source, which is $(6.3\pm 0.8) \times 10^{20}\,\rm cm^{-2}$ based on the extinction/absorption maps presented in \citet{Doroshenko2024}\footnote{\url{http://astro.uni-tuebingen.de/nh3d/nhtool}}. 

The fitting parameters are crucially depended on the obtained absorption parameter~$\nh$~\citep[see e.g.][]{Suleimanov_etal2024}. Consequently, we adopted $\nh=(5.5\pm1)\times 10^{20}\,\rm cm^{-2}$, as obtained by \citet{Gaensicke_etal1998} from the analysis of the UV~spectra of the source, and we use this value further. It should be noted that this issue also arises when fitting soft X-ray spectra of other~SSSs with atmosphere models~\citep[see e.g.][]{Rauch_etal2010, Suleimanov_etal2024, Tavleev_etal2024}. 

\subsection{WD mass estimation}

\begin{figure}
\centering
	\center{\includegraphics[width=0.95\linewidth]{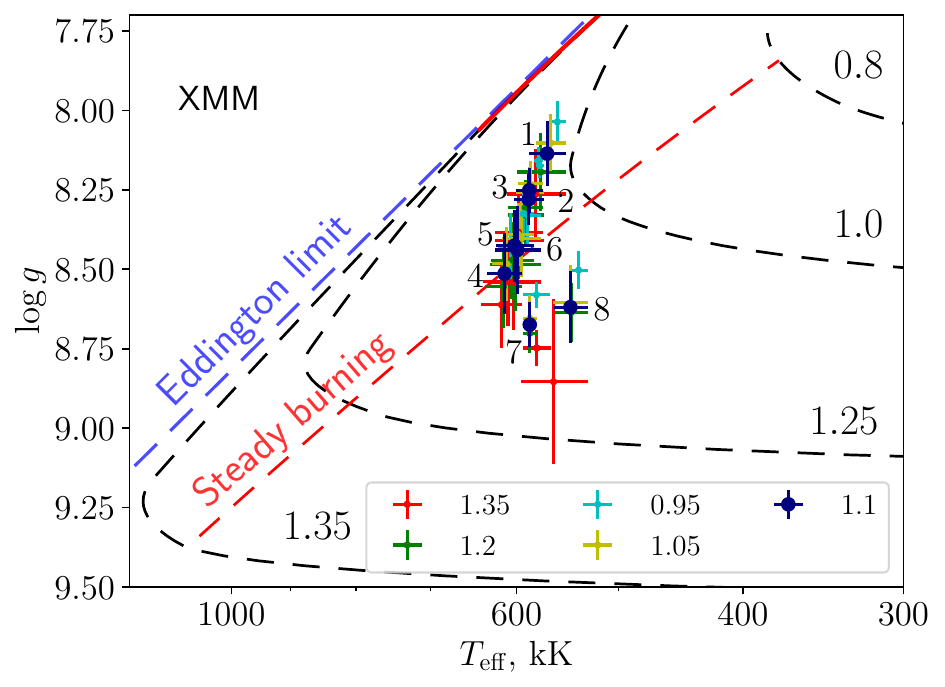}}
 	\center{\includegraphics[width=0.95\linewidth]{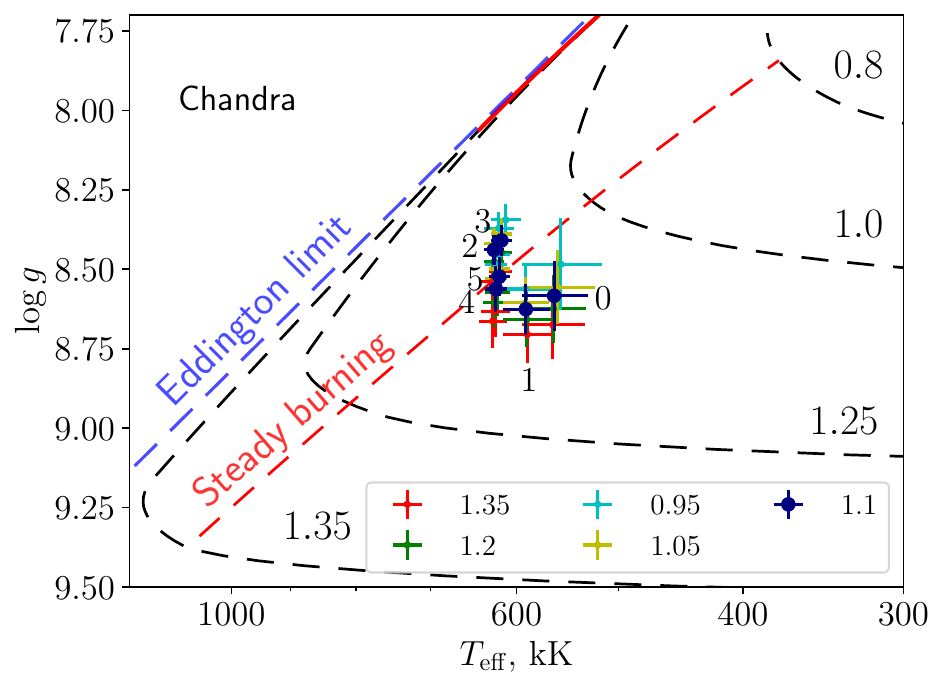}}
    \caption{Positions of the source in the $\Teff - \log\varg$ plane according to different \xmm~(upper panel) and \chandra~(lower panel) observations~(see Tabs.~\ref{tab:fit_chandra}~and~\ref{tab:fit_xmm}). The WD mass was fixed, and different colours indicate different masses from~$0.95$ to~$1.35\, M_{\sun}$. For clarity, only a portion of the mass range is shown. The fits with $M=1.1\,M_{\sun}$ are additionally indicated by the increased marker size. Model dependencies for various WD~masses, taken from~\citep{Nomoto_etal2007}, are shown by black dashed curves. The numbers at the curves indicate WD~masses (in solar masses). The lower boundary of the stable thermonuclear burning band is shown by the dashed red line. The Eddington limit for solar H/He abundances is shown by the blue dashed line. The numbers denote the spectrum number~(see Sect.~\ref{sect:Observations}).}
    \label{fig:glogT}
\end{figure}

\begin{figure}
    \centering
    \includegraphics[width=0.94\columnwidth]{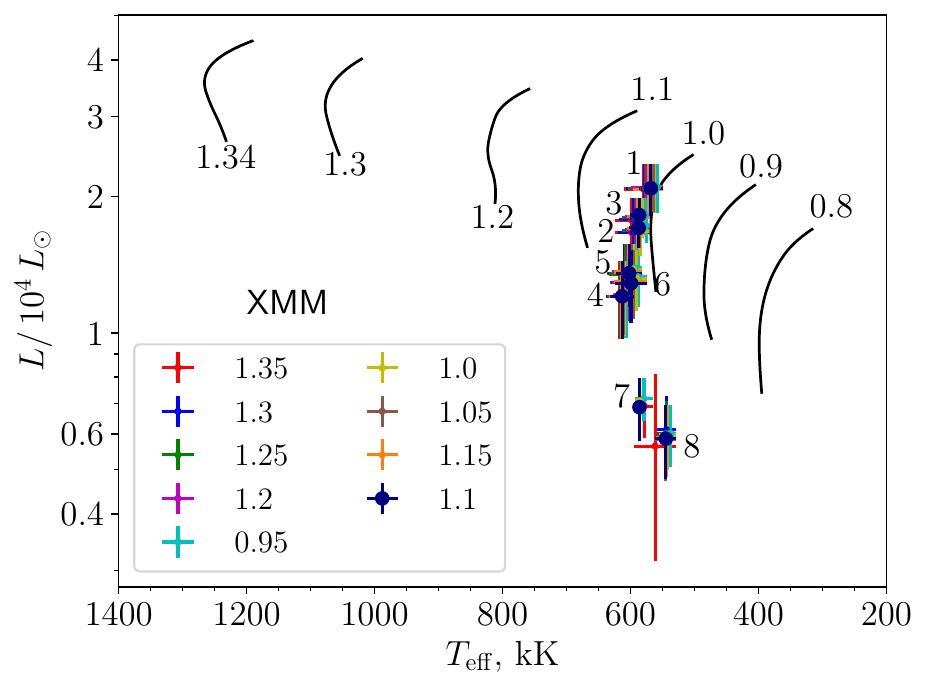}
    \includegraphics[width=0.94\columnwidth]{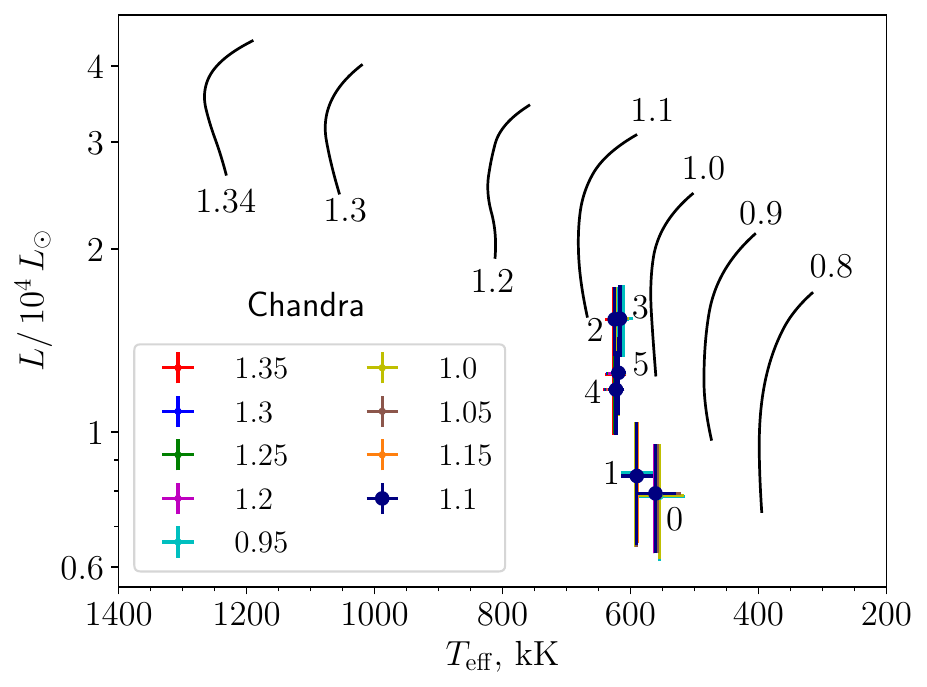}
    \caption{Positions of the source in the $\Teff - L$ plane according to different \xmm~(upper panel) and \chandra~(lower panel) observations. The WD~mass was fixed, and different colours indicate different masses from~$0.95$ to~$1.35\, M_{\sun}$. The fits with $M=1.1\,M_{\sun}$ are additionally indicated by the increased marker size. Model dependencies were taken from~\citet{Wolf_etal2013}. Only the model curves with steady-state thermonuclear burning are shown. The numbers denote the spectrum number~(see Sect.~\ref{sect:Observations}).}
    \label{fig:LT} 
\end{figure}

The obtained values of the WD~mass vary significantly (from~$0.8$ to~$1.38\,\rm M_{\sun}$ for the different observations), if we consider the mass as a free parameter.
Moreover, the fits with higher masses yield a WD~radius close to the lower limit for WD~radii. In fact, this lower limit is about~$3000\,\rm km$, which appears to be an incorrect radius for the~WD in~RXJ0513.
This means that we cannot accurately determine the WD~mass using spectral fitting alone. On the other hand, the obtained~$\log\varg$ and~$\Teff$ values do not vary significantly. Therefore, we apply an alternative approach to estimate the WD~mass, based on comparing the obtained~$\log\varg$,~$\Teff$, and bolometric luminosity~$L$ for each observation with model predictions for WDs with thermonuclear burning on their surface. This approach has been successfully used previously~\citep{Suleimanov_etal2024,Tavleev_etal2024}.

We fitted all the spectra using nine fixed WD~mass values from~$0.95$ to~$1.35\,\rm M_{\sun}$ in steps of~$0.05\,\rm M_{\sun}$ and then compared the obtained results with theoretical predictions. 
Namely, we put the obtained fit values on the theoretical~$\Teff-\log\varg$ and $\Teff-L$~dependencies, computed for different WD~masses with hydrogen-rich envelopes with thermonuclear burning by \citet{Nomoto_etal2007} and \citet{Wolf_etal2013}. The results obtained are shown in Figs.~\ref{fig:glogT}~and~\ref{fig:LT}.
All values obtained from the fits lie within the mass range of~$1.0-1.15\, M_{\sun}$. Eventually, we consider $M=1.1\, M_{\sun}$ as the correct WD~mass (with uncertainty about~$0.1\,M_{\sun}$), and the parameters corresponding to this WD~mass are indicated in Figs.~\ref{fig:glogT}~and~\ref{fig:LT} by an increased marker size. It should be noted that in the previous paper \citep{Suleimanov_etal2024} we estimated the WD~mass in the source as~${>}\,1.15\, M_{\sun}$, although only one \chandra~observation~(C3) was used.

\subsection{Evolution of RXJ0513}

\begin{figure*}
    \center{\includegraphics[width=1.0\linewidth]{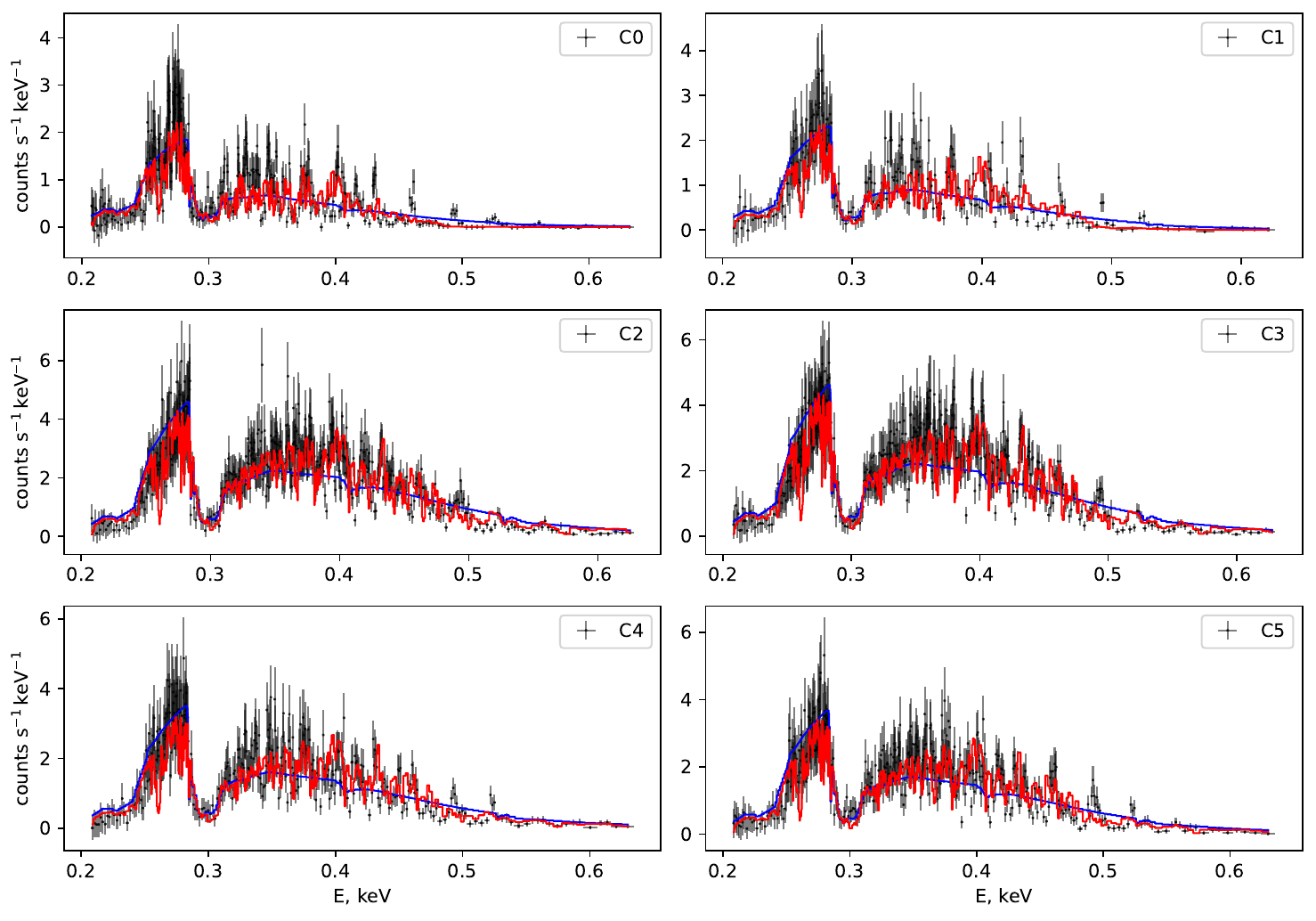}}
    \caption{\chandra~spectra of~RXJ0513 with the best-fit LTE models~(in red). The absorbed blackbody models are also shown~(in blue). The obtained model parameters are listed in Table~\ref{tab:fit_chandra}. The hydrogen column density is fixed at~$\nh=5.5\times 10^{20}\,\rm cm^{-2}$, and the mass is set at $M=1.1\, M_\sun$.}
    \label{fig:spectra_all_chandra}
\end{figure*}

\begin{figure*}
    \center{\includegraphics[width=1.0\linewidth]{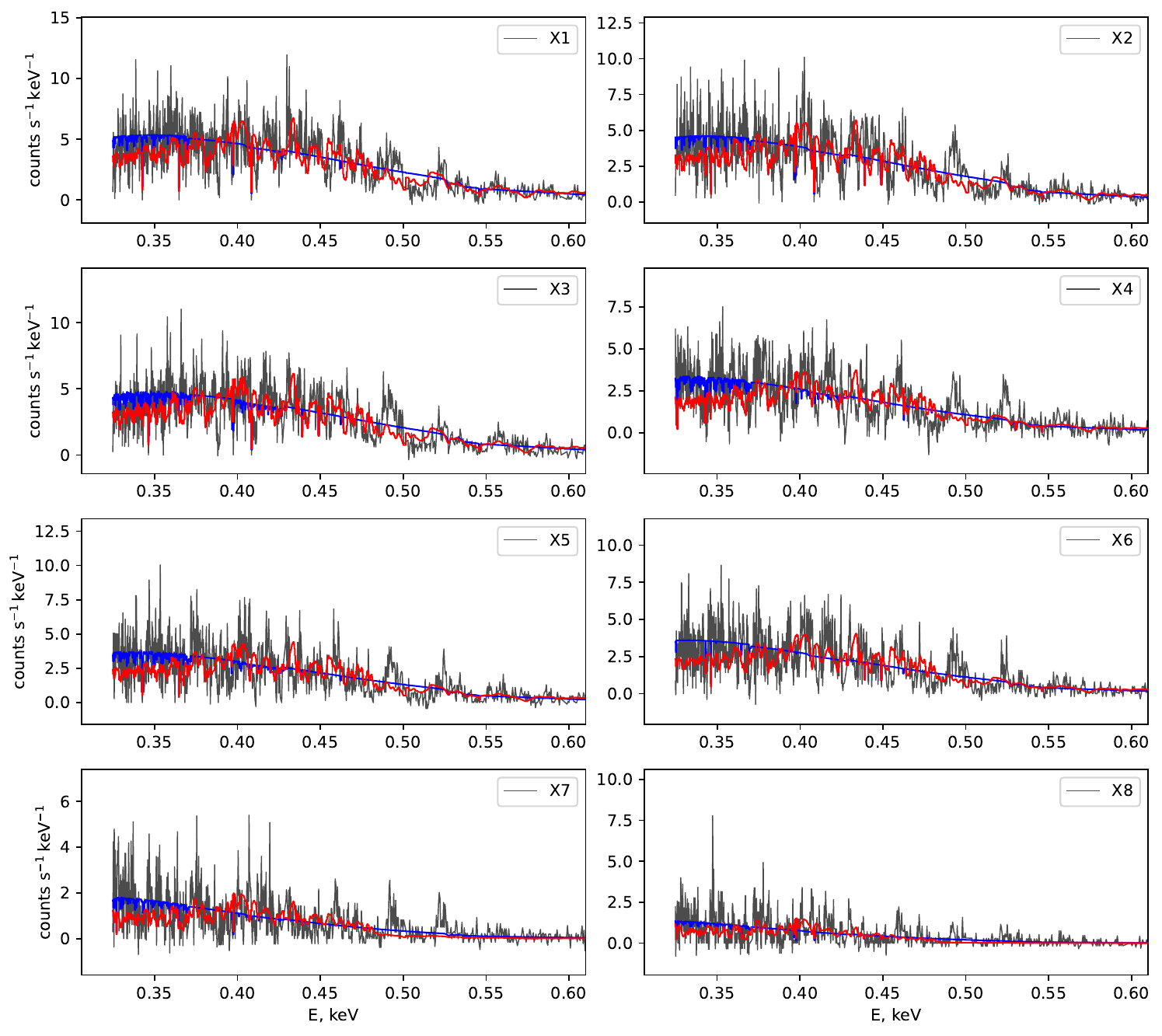}}
    \caption{Same as in Fig.~\ref{fig:spectra_all_chandra}, but for \xmm~observations, RGS1~instrument. Only every second count in the observational spectra is shown for clarity.}
    \label{fig:spectra_all_xmm}
\end{figure*}

\begin{table*}
\begin{center}
    \caption{Spectral parameters of the LTE and blackbody model fit for the \chandra~spectra.}
    \label{tab:fit_chandra}
    \begin{tabular}{ccccc|ccc}
    \hline \hline \\ [-0.7em]
     & $\Teff$ & $R^a$ & $\log\varg$ & $L$ & $\Teff^{\rm bb}$ & $R^a_{\rm bb}$ & $L_{\rm bb}$ \\ [0.3em]
     & kK & km & & $10^{37}\,\rm erg\,s^{-1}$ & kK & km & $10^{37}\,\rm erg\,s^{-1}$ \\ [0.3em]
    \hline \\ [-0.7em] 
    C0 & $561\pm 33$ & $6297\pm 660$ & $8.58\pm 0.11$ & $3.0\pm 0.6$ & $559\pm 24$ & $7776\pm 984$ & $4.3\pm 1.2$ \\ [0.3em]
    C1 & $590\pm 25$ & $5993\pm 550$ & $8.63\pm 0.08$ & $3.2\pm 0.7$ & $589\pm 24$ & $7159\pm 970$ & $4.4\pm 1.4$ \\ [0.3em]
    C2 & $625\pm 11$ & $7351\pm 401$ & $8.44\pm 0.05$ & $5.9\pm 0.8$ & $706\pm 18$ & $6121\pm 417$ & $6.7\pm 1.1$ \\ [0.3em]
    C3 & $616\pm 11$ & $7569\pm 432$ & $8.41\pm 0.05$ & $5.9\pm 0.8$ & $702\pm 25$ & $6191\pm 519$ & $6.8\pm 1.2$ \\ [0.3em]
    C4 & $623\pm 13$ & $6370\pm 544$ & $8.56\pm 0.07$ & $4.5\pm 0.7$ & $662\pm 17$ & $6322\pm 533$ & $5.5\pm 1.1$ \\ [0.3em]
    C5 & $619\pm 12$ & $6683\pm 512$ & $8.52\pm 0.06$ & $4.8\pm 0.7$ & $675\pm 23$ & $6114\pm 519$ & $5.7\pm 1.1$ \\ [0.3em]
    \hline
    \end{tabular}
\end{center}
{\small Notes: (a)~-- the distance to the~LMC is assumed to be~$50\,\rm kpc$~\citep{Pietrzynski_etal2019}; 
$\nh=5.5\times 10^{20}\,\rm cm^{-2}$ and~$M=1.1\, M_{\sun}$ are fixed for all of the fits.}
\end{table*}

\begin{table*}
\begin{center}
    \caption{Spectral parameters of the LTE and blackbody model fit for the \xmm~spectra.}
    \label{tab:fit_xmm}
    \begin{tabular}{ccccc|ccc}
    \hline \hline \\ [-0.7em]
     & $\Teff$ & $R^a$ & $\log\varg$ & $L$ & $\Teff^{\rm bb}$ & $R^a_{\rm bb}$ & $L_{\rm bb}$ \\ [0.3em]
     & kK & km & & $10^{37}\,\rm erg\,s^{-1}$ & kK & km & $10^{37}\,\rm erg\,s^{-1}$ \\ [0.3em]
    \hline \\ [-0.7em] 
    X1 & $568\pm 19$ & $10324\pm 1150$ & $8.14\pm 0.10$ & $8.0\pm 1.1$ & $553\pm 16$ & $17765\pm 2227$ & $21.3\pm 3.5$ \\ [0.3em]
    X2 & $587\pm 16$ & $8775\pm 797$ & $8.28\pm 0.08$ & $6.5\pm 0.6$ & $537\pm 16$ & $18575\pm 2615$ & $20.7\pm 4.0$ \\ [0.3em]
    X3 & $586\pm 14$ & $9065\pm 727$ & $8.25\pm 0.07$ & $7.0\pm 0.6$ & $553\pm 16$ & $16961\pm 2178$ & $19.4\pm 3.5$ \\ [0.3em]
    X4 & $613\pm 19$ & $6748\pm 980$ & $8.51\pm 0.13$ & $4.6\pm 0.9$ & $519\pm 8$  & $17984\pm 1512$ & $17.0\pm 2.8$ \\ [0.3em]
    X5 & $602\pm 21$ & $7417\pm 923$ & $8.43\pm 0.11$ & $5.2\pm 0.8$ & $522\pm 18$ & $18680\pm 3012$ & $18.7\pm 4.3$ \\ [0.3em]
    X6 & $599\pm 25$ & $7348\pm 1135$ & $8.44\pm 0.14$ & $4.9\pm 0.9$ & $502\pm 19$ & $21202\pm 3744$ & $20.7\pm 5.2$ \\ [0.3em]
    X7 & $586\pm 7$ & $5562\pm 441$ & $8.67\pm 0.07$ & $2.6\pm 0.4$ & $439\pm 24$ & $27610\pm 8135$ & $20.7\pm 9.0$ \\ [0.3em]
    X8 & $545\pm 17$ & $6007\pm 685$ & $8.62\pm 0.11$ & $2.2\pm 0.4$ & $413\pm 31$ & $29897\pm 9880$ & $18.7\pm 8.5$ \\ [0.3em]
    \hline
    \end{tabular}
\end{center}
{\small Notes: (a)~-- the distance to the~LMC is assumed to be~$50\,\rm kpc$~\citep{Pietrzynski_etal2019}; 
$\nh=5.5\times 10^{20}\,\rm cm^{-2}$ and~$M=1.1\, M_{\sun}$ are fixed for all of the fits.}
\end{table*}

Tabs.~\ref{tab:fit_chandra}~and~\ref{tab:fit_xmm} present the final results of our analysis, while Figs.~\ref{fig:spectra_all_chandra}~and~\ref{fig:spectra_all_xmm} show the comparison between the observed and model spectra. The temperatures~$\Teff$ derived from our LTE~models are higher than those found by \citet{McGowan_etal2005}, who used a blackbody model to fit the EPIC-PN spectra of~RXJ0513. In contrast, the radii obtained are much smaller than the values determined by the blackbody fits. It should be noted that we fixed the hydrogen column density~$\nh$ at~$5.5\times 10^{20}$, whereas \citet{McGowan_etal2005} used a value of~$6.2\times 10^{20}$. 

\begin{figure*}
\centering
    \center{\includegraphics[width=0.9\linewidth]{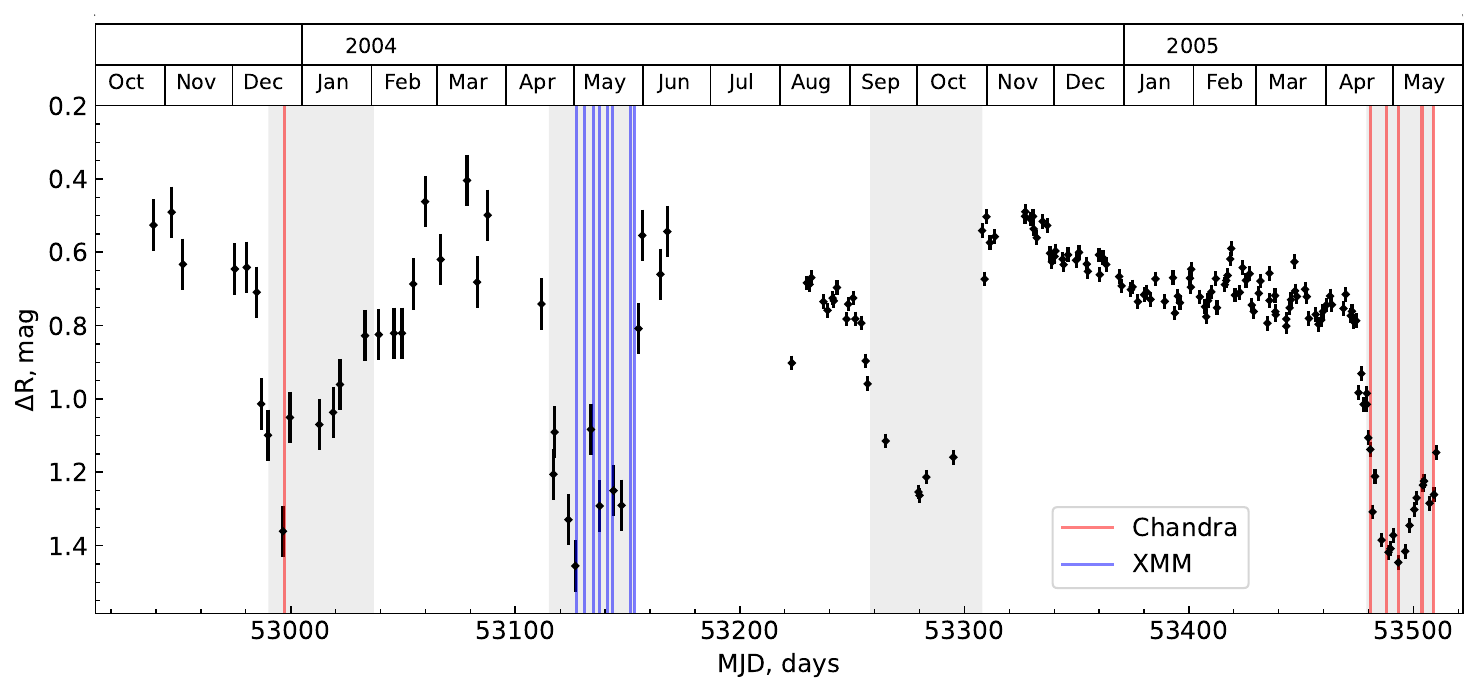}}
    \caption{Optical monitoring of RX~J$0513.9-6951$, differential magnitudes in the R-filter, obtained by \citet{McGowan_etal2005} and \citet{Burwitz_etal2008}. The times of the observed optical low states are shaded in grey. The dates of X-ray \chandra~and \xmm~observations are shown in red and blue.}
    \label{fig:light_curve}
\end{figure*}

\begin{figure*}
\centering
    \begin{minipage}{0.49\linewidth}
    \center{\includegraphics[width=1.0\linewidth]{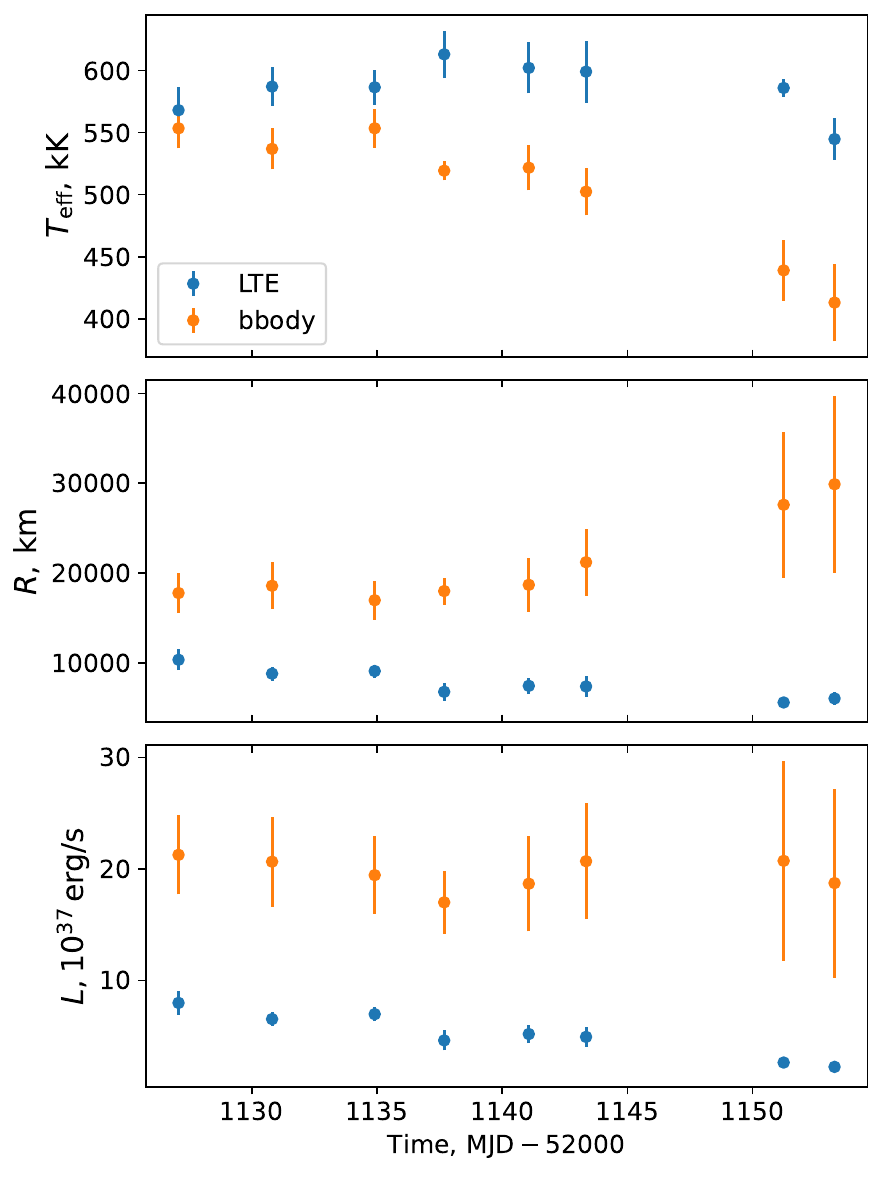}}
    \end{minipage}
    \begin{minipage}{0.49\linewidth}
    \center{\includegraphics[width=1.0\linewidth]{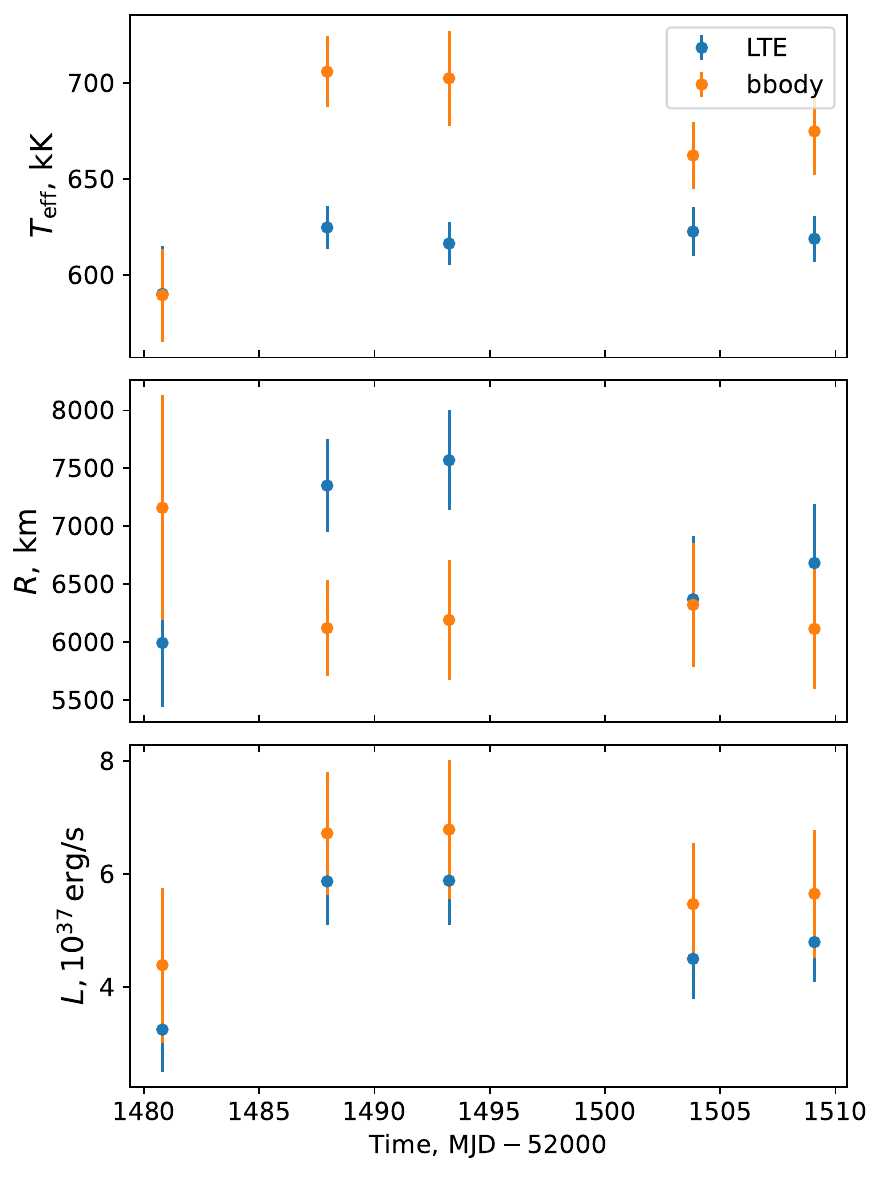}}
    \end{minipage}
    \caption{Evolution of temperature, radius, and luminosity of the~WD in~RXJ0513 over time during the \xmm~(left panels) and \chandra~(right panels) observations. Results are shown for both the model atmosphere fit and the blackbody fit. The WD~mass is fixed, $M = 1.1\,M_{\sun}$.}
    \label{fig:Teff_R_L_MJD}
\end{figure*}

We can now examine the X-ray evolution of~RXJ0513 and compare the results with the optical brightness of the system. The optical light curve is presented in Fig.~\ref{fig:light_curve}, while for the optical data description we refer to \citet{McGowan_etal2005} and \citet{Burwitz_etal2008}.
The observational dates of the investigated X-ray spectra are also marked in the figure. They cover three optical low states, corresponding to the~C0,~X1-8, and~C1-5 observational sequences. Spectra X1-6 correspond to the optical low state of the source, while~X7-8 depict an increase in optical flux, marking the transition to the optical high state. Similarly, during spectrum~C1, the transition to the optical low state begins; the source remains optically faint during C2-3, and the transition to the optical high state occurs during C4-5.
 
The changes in the~WD parameters, obtained from fitting the X-ray spectra, with time are shown in Fig.~\ref{fig:Teff_R_L_MJD}.
For the \xmm~spectra, it is evident that initially the temperature increases, accompanied by a decrease in both the radius and luminosity. Later, the temperature starts to decrease, and during the transition from low to high optical state, temperature, radius, and luminosity all approach their minimum values.
For the \chandra~spectra, temperature, radius, and luminosity all increase over time during the transition from the high to low optical state. Later, during the transition from the low to high optical state, the temperature remains stable, while radius and luminosity decrease to values comparable the initial ones.

The results of the blackbody fitting are also shown in Fig.~\ref{fig:Teff_R_L_MJD}. The blackbody fit parameters for the \chandra~observations roughly correspond to those from the model atmosphere fits with slightly overestimated bolometric luminosities, although the radii differ rather randomly.
In contrast, the blackbody fit parameters from the \xmm~observations differ more significantly from the atmosphere model fits, yielding radii and luminosities that are several times larger. Overall, it can be stated that the blackbody fits are unsuitable for analysing the \xmm~spectra of~SSSs.

\begin{figure*}
\centering
    \begin{minipage}{0.49\linewidth}
    \center{\includegraphics[width=1.0\linewidth]{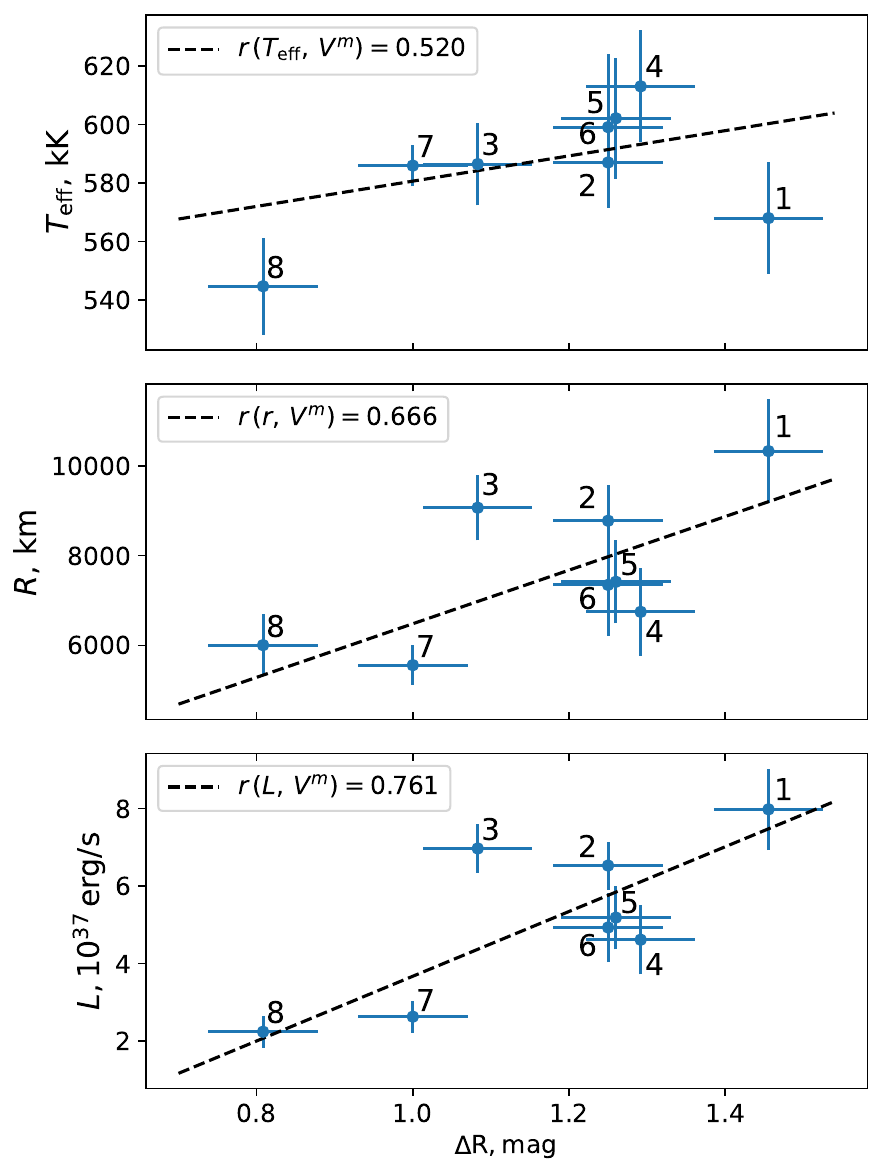}}
    \end{minipage}
    \begin{minipage}{0.49\linewidth}
    \center{\includegraphics[width=1.0\linewidth]{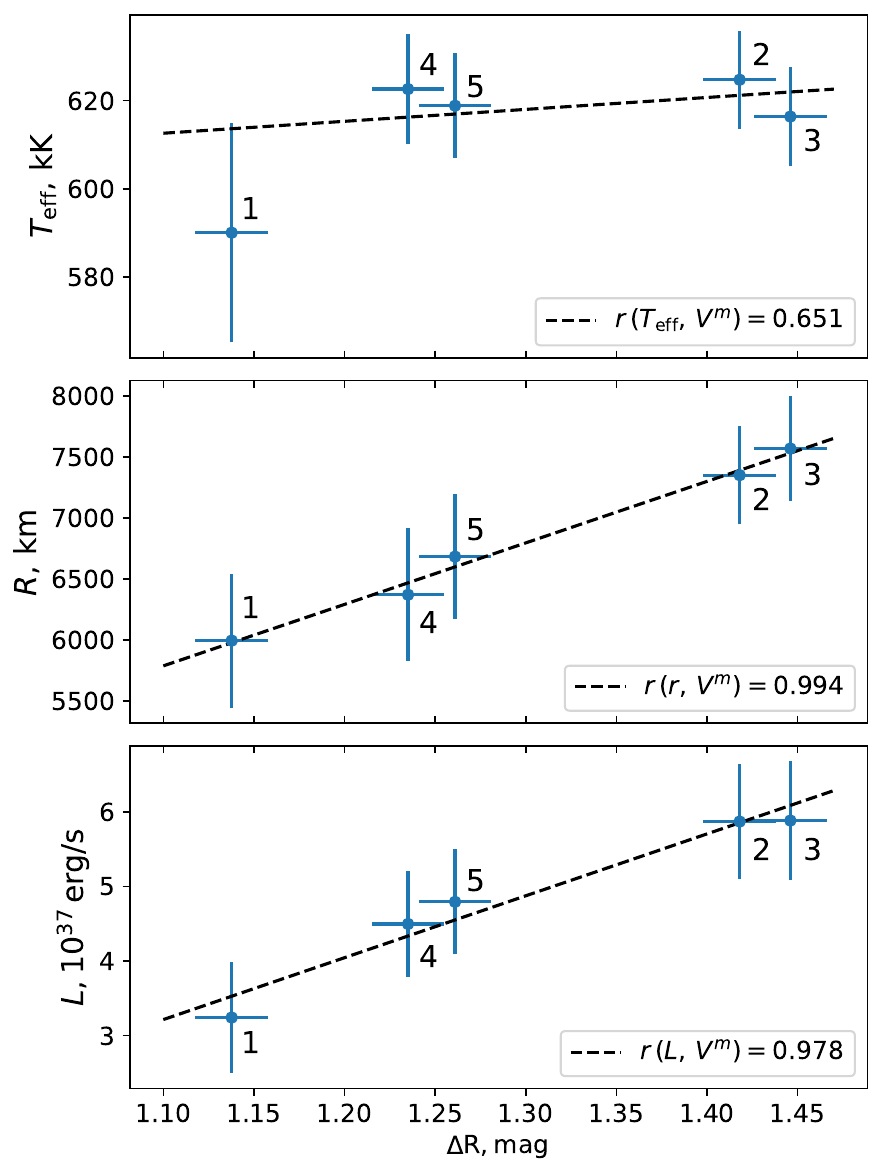}}
    \end{minipage}
    \caption{Temperature, radius, and bolometric luminosity of the~WD in~RXJ0513 versus the optical brightness of the source for the \xmm~(left panels) and \chandra~(right panels) spectra. The numbers denote the spectrum number~(see Sect.~\ref{sect:Observations}). The linear approximation as well as the Pearson correlation coefficients between the optical flux and WD~parameters are also shown.}
    \label{fig:pars_OPflux}
\end{figure*}

Fig.~\ref{fig:pars_OPflux} illustrates the relationships between the WD~parameters, obtained using the model atmosphere spectra, and the optical brightness of RXJ0513. One can clearly see (especially for the \chandra~spectra, where the Pearson correlation coefficient is close to unity) the correlation between radius (and bolometric luminosity) and the optical brightness of RXJ0513~-- the larger the~WD (and the bolometric luminosity) the lower the optical brightness.

By comparing the position of the source on the $\Teff-\log\varg$ diagram, obtained during the different \xmm~observations, with theoretical tracks~(see Fig.~\ref{fig:glogT}) we conclude that when~RXJ0513 is optically bright~(in observations~X7-X8), it is situated below the steady-burning strip. This means that the nuclear burning on the~WD's surface stops, and the~WD begins to cool, when its optical brightness is high. And vice versa, the source undergoes steady burning when it is faint in the optical band.

The same conclusion is correct for \chandra~observations. The source is located in the stable-burning strip when it is optically bright~(observations C2,~C3 and~C0). During the other observations,~RXJ513 is optically fainter and it is situated above or at the boundary of the steady-burning strip~(see Fig.~\ref{fig:glogT}).

As shown in Fig.~\ref{fig:glogT}, the higher the fixed WD~mass, the higher the surface gravity and the effective temperature of the~WD. This means that the positions of more massive~WDs move towards below the steady-burning region in the~$\Teff - \log\varg$ plane. 
While for the \xmm~spectra our conclusions about steadiness of the source on this plane hold for the entire mass range considered, all the positions obtained using the \chandra~observations lie below the steady-burning limit when $M\geq1.3\,M_{\sun}$.

Modelling the \xmm~spectra with blackbody and fixed~$\nh$ shows a clear decrease in~$\Teff$ and increase in radius, as shown in Fig.~\ref{fig:Teff_R_L_MJD} (left panels, in orange) and Tab.~\ref{tab:fit_xmm}. \citet{McGowan_etal2005} obtained the similar result for the EPIC-PN spectra of~\xmm.
This behaviour of the source does not correspond to the results obtained using the model atmosphere spectra. Moreover, the obtained radii and luminosities are significantly larger even at a fixed~$\nh$. The blackbody fit for the \chandra~observations aligns more closely with the model atmosphere results.
Fixing the column density at~$\nh=6.2\times10^{20}$, the value found by \citet{McGowan_etal2005}, does not change the result significantly. From a statistical point of view, this $\nh$~value is less likely. The theoretical tracks of~WD for all spectra shift slightly towards lower temperatures and gravities (upwards and to the right in Fig.~\ref{fig:glogT}), but the overall stability characteristics remain almost unchanged, as does the evolution over time.

\section{Discussion}
\label{sect:Discussion}

The observed anti-correlation between optical and soft X-ray fluxes in~RXJ0513 is usually explained by the so-called contraction model~\citep[e.g.][]{Reinsch_etal1996, Reinsch_etal2000, Southwell_etal1996, Hachisu_etal2003a, Hachisu_etal2003b}. According to this model, the mass accretion rate~$\dot{M}$ is relatively low during the X-ray bright state, and the nuclear burning rate in the envelope is close to the steady-burning boundary. This state of the source corresponds to an almost minimal radius of the~WD~photosphere, with the highest effective temperature and a relatively low bolometric luminosity. When~$\dot{M}$ increases, the~WD inflates and the bolometric luminosity slightly increases, while the effective temperature drops. The decrease in effective temperature suggests that the peak of the spectral energy distribution shifts towards lower photon energies, and a smaller fraction of the emitted spectrum falls within the observed soft X-ray band. Therefore, the increase in the WD~photosphere radius at an almost constant bolometric luminosity leads to a decrease in the observed soft X-ray flux.
Thus, within the framework of the contraction model, the low X-ray flux corresponds to a high mass accretion rate, a large WD~radius and a high bolometric luminosity. Conversely, the bright X-ray state is associated with a hotter~WD with a relatively small photospheric radius and a lower bolometric luminosity.

In the considered model the optical flux is produced by accretion disc illumination and reprocessing of hard radiation into the optical band. A colder~WD with an expanded photosphere illuminates the accretion disc more
significantly, meaning that the faint X-ray state should correspond to a high optical state.
Conversely, a hotter~WD with a deflated photosphere illuminates the accretion disc less effectively, so the low optical state corresponds to the bright X-ray state, leading to the observed anti-correlation between optical and soft X-ray fluxes from the source. However, our results contradict this model. We show in Fig.~\ref{fig:pars_OPflux} that the bright optical state of~RXJ0513 corresponds to a low X-ray luminosity, during which the source lies below the stable-burning strip on the $\Teff - \log\varg$ plane~(see Fig.~\ref{fig:glogT}).

The found contradiction could be explained if we take into account the effects of thermalisation of hard radiation and its re-emission in the optical part of the spectrum. The disc photosphere itself transforms the external soft X-ray radiation into the optical band very inefficiently. The reprocessing efficiency for soft X-rays and far~UV radiation in irradiated accretion disc photospheres in~SSSs is low~\citep[$\eta\sim 0.05-0.1,\, F_{\rm opt}=\eta\, F_{\rm soft\, X-ray}$, see e.g.][]{Suleimanov_etal1999}.
The opacity of the upper atmospheric layers in soft X-ray is so high that the irradiated flux is absorbed before reaching the optical flux formation depth. The absorbed flux is then re-radiated mainly in far UV~spectral lines.

\begin{figure}
\centering
	\center{\includegraphics[width=1.0\linewidth,trim=0.5cm 1.5cm 0.53cm 0.5cm, clip]{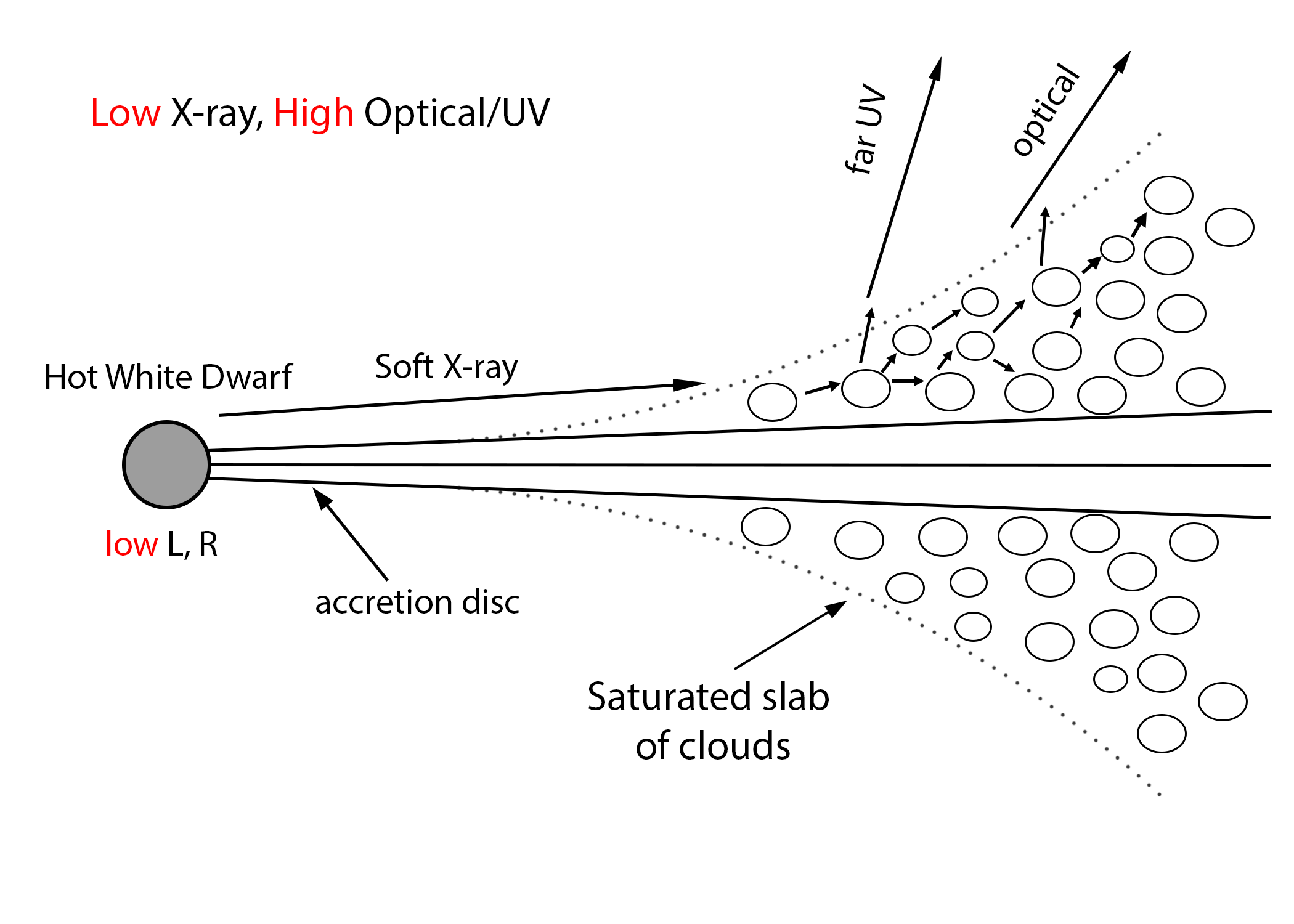}}
 	\center{\includegraphics[width=1.0\linewidth,trim=0.35cm 2.6cm 0.5cm 1.32cm, clip]{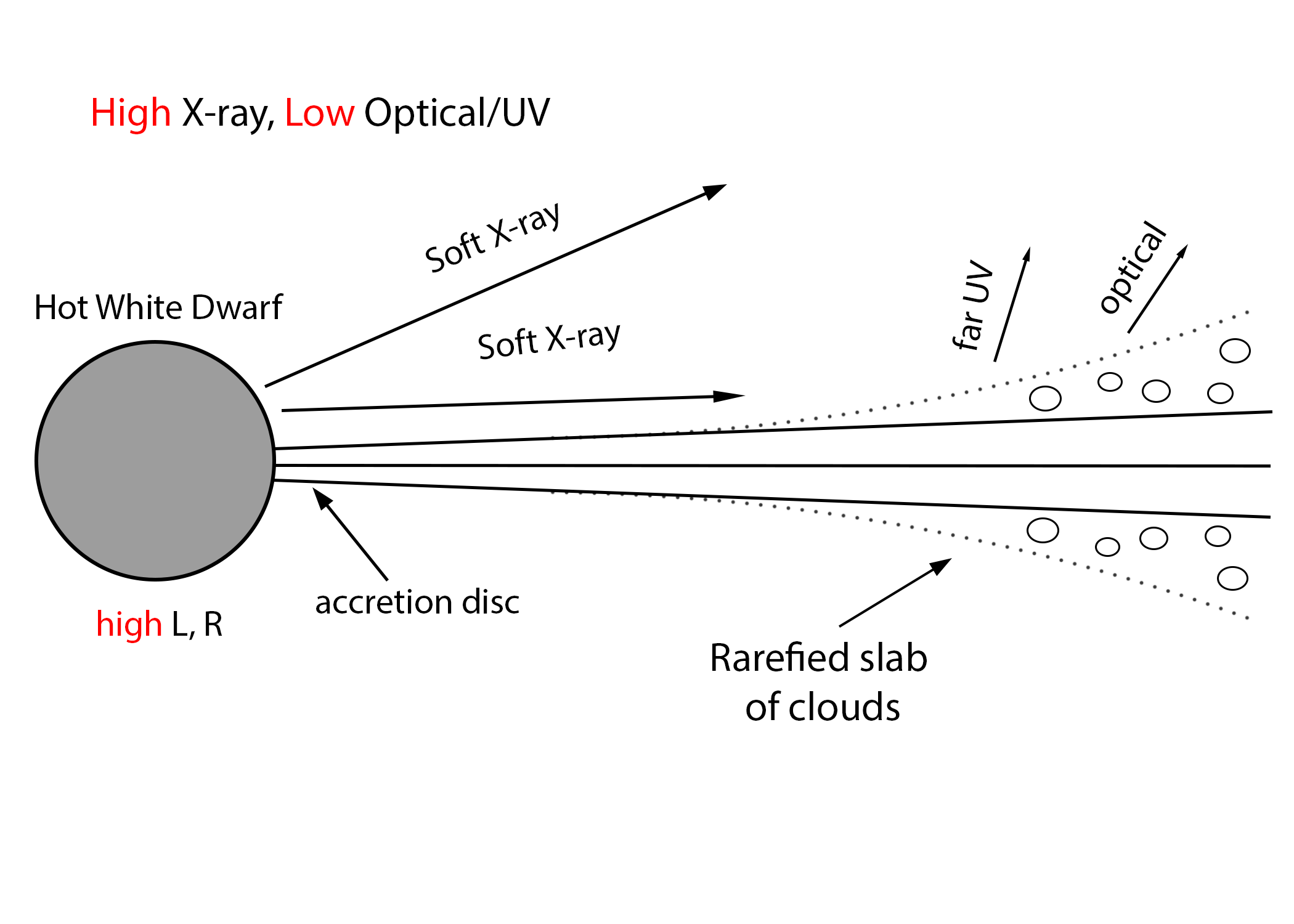}}
    \caption{Qualitative picture of the proposed high and low optical states. Shown are the~WD, the accretion disc, and blobs or clouds above the disc. Upper panel~-- high optical state; the low luminous~WD is compact, and its X-ray emission is effectively reprocessed into optical light due to multiple scattering between clouds. Lower panel~-- low optical state; the luminous~WD with an expanded photosphere evaporates most of clouds decreasing the reprocessing efficiency and reducing the optical flux.}
    \label{fig:model_image}
\end{figure}

Reprocessing efficiency can increase to acceptable values of~$\eta\sim 0.3-0.5$ if we assume a system of optically thick~(in soft X-ray) clouds above the disc. In this case, the total reprocessing efficiency rises due to multiple scattering of soft X-ray and far~UV radiation between clouds~\citep{Suleimanov_etal2003}. Figure~\ref{fig:model_image} presents a qualitative illustration of the proposed model. The value of~$\eta$ depends on the effective optical depth of the cloud slab~\citep{Suleimanov_etal2003},
\begin{equation}
       \tau_{\rm eff} \approx \pi \,R_{\rm cl}^2\,N_{\rm cl} \,L,    
\end{equation}
where $R_{\rm cl}$ is the cloud size, $N_{\rm cl}$ is the cloud number density, and $L$~is the geometrical thickness of the cloud slab. The reprocessing efficiency reaches the maximum at~$\tau_{\rm eff} \approx 1-10$.

The nature of these hypothetical clouds is not known. \citet{Schandl_etal1997} proposed that the clouds or blobs could be a `spray' resulting from the interaction between the accreting gas stream and the outer boundary of the accretion disc. However, clouds formed due to the thermal instability of the matter illuminated by photoionising radiation are more likely~\citep[see e.g.][]{2002ApJ...581.1297J}. Thus, the changes in the optical flux of~RXJ0513 could be explained by assuming that thermal instability conditions depend on the state of the WD~photosphere. Namely, a relatively small and low luminous WD~photosphere provides better conditions for cloud formation. Consequently, the effective optical thickness of the cloud slab increases and we observe the bright optical state of the source, accompanied by the faint X-ray flux. Moreover, the cloud slab might even become so geometrically thick that it obscures the entire X-ray source. And vice versa, a luminous~WD with an expanded photosphere evaporates the clouds, reducing~$\tau_{\rm eff}$, and thereby decreasing reprocessing efficiency. Both hypothetical states are illustrated in Fig.~\ref{fig:model_image}.

\section{Conclusions}
\label{sect:Conclusions}

In this work, we performed a spectral analysis of the supersoft X-ray source RX~J$0513.9-6951$, which was observed in X-rays by the \chandra~and \xmm~telescopes during its optically low states, when the source exhibits maximum X-ray brightness.
To describe the spectra we used a grid of the high-gravity hot LTE~model atmospheres, calculated by~\citet{Suleimanov_etal2024}. The grid was computed for~$\Teff$ in the range of~$100-1000\rm\, kK$ in steps of~$25\,\rm kK$. The $\dlg=\log\varg - \log\varg_{\rm Edd}$~parameter has eight values: $0.1, 0.2, 0.4, 0.6, 1.0, 1.4, 1.8$~and~$2.2$. We adopted the LMC~composition, with the hydrogen-helium mix set to solar and the abundance of heavy elements set to half-solar~$(A = 0.5)$. 

We performed two joint fits of the \xmm~and~\chandra~spectra with different values of a common WD~mass parameter and then compared the results with theoretical $\Teff - \log\varg$ and $\Teff - L$ model curves. The fit positions correspond to $M=1.1\pm 0.1\,M_\sun$, which we adopt as the obtained WD~mass.

Our analysis reveal the evolution of the~WD in RXJ0513: the photospheric radius of the~WD and its bolometric luminosity increase as the optical flux decreases, and vice versa. This conclusion is based on studying the source's position on the~$\Teff- \log\varg$ plane, where the source lies below the stable-burning strip when it is optically bright. As its optical brightness decreases, RXJ0513 shifts towards the stable-burning strip. We also find the correlation between the photospheric WD~radius and the source magnitude in R-band, as well as between the bolometric luminosity and R-band magnitude.
These results contradict the predictions of the contraction model, which is commonly used to explain the observed anti-correlation between soft X-ray and optical fluxes~\citep[see e.g.][]{Reinsch_etal1996, Reinsch_etal2000, Southwell_etal1996}. This model predicts the opposite correlation between the WD~photospheric radius and its optical brightness.

We propose an alternative model of~RXJ0513 periodicity based on the reprocessing model suggested by \citet{Suleimanov_etal2003}. In this model, efficient reprocessing of soft X-ray flux illuminating the accretion disc is achieved through multiple scattering of hard radiation in a system of optically thick~(in the soft X-ray band) gaseous blobs/clouds above the accretion disc, embedded in a disc wind or corona. These clouds may arise due to thermal instability in the matter above the disc, irradiated by far UV/soft X-ray flux.

We suggest that the cloud system becomes highly saturated when the~WD has relatively low luminosity and a small radius, comparable to that of a cold~WD. In this case, a cloud slab provides reprocessing efficiency high enough to transform the soft X-ray/far UV flux into the optical band. This picture is proposed to represent the~RXJ0513 state with low X-ray flux and high optical brightness. Alternatively, a luminous~WD with an expanded photosphere is likely less favourable for clouds formation. The cloud slab becomes rarefied, and the reprocessing efficiency decreases. This picture corresponds to the source's low optical state, accompanied by high X-ray flux.

This suggested toy model qualitatively explains the observed properties of~RXJ0513 and may open a window for observational investigations of thermal instability in plasma irradiated by hard UV/soft X-ray flux.

\begin{acknowledgements}
    {This work was supported by the \emph{Deut\-sche For\-schungs\-ge\-mein\-schaft\/} under grants WE1312/56--1 (AT) and WE1312/59--1 (VFS).}
\end{acknowledgements}

\bibliographystyle{aa} 
\bibliography{references} 

\end{document}